# Radiation-induced D/H Exchange Rate Constants in Aliphatics Embedded in Water Ice

Danna Qasim[1,2,3], Reggie L. Hudson[1], and Christopher K. Materese[1]
[1] Astrochemistry Laboratory, NASA Goddard Space Flight Center Greenbelt, MD 20771, USA; christopher.k.materese@nasa.gov
[2] Department of Physics and Astronomy, Howard University, Washington, DC, 20059, USA
[3] Center for Research and Exploration in Space Science and Technology, NASA/GSFC Greenbelt, MD 20771, USA

## Abstract

Gas-phase and solid-state chemistry in low-temperature interstellar clouds and cores leads to a D/H enhancement in interstellar ices, which is eventually inherited by comets, meteorites, and even planetary satellites. Hence, the D/H ratio has been widely used as a tracer for the origins of extraterrestrial chemistry. However, the D/H ratio can also be influenced by cosmic rays, which are ubiquitous and can penetrate even dense interstellar molecular cores. The effects of such high-energy radiation on deuterium fractionation have not been studied in a quantitative manner. In this study, we present rate constants for radiation-induced D-to-H exchange for fully deuterated small (1–2 C) hydrocarbons embedded in $H_2O$ ice at 20 K and H-to-D exchange for the protiated forms of these molecules in $D_2O$ ice at 20 K. We observed larger rate constants for H-to-D exchange in the $D_2O$ ice versus D-to-H exchange in $H_2O$ ice, which we have attributed to the greater bond strength of C–D versus C–H. We find that the H-to-D exchange rate constants are smaller for protiated methane than ethane, in agreement with bond energies from the literature. We are unable to obtain rate constants for the unsaturated and reactive hydrocarbons ethylene and acetylene. Interpretation of the rate constants suggest that D/H exchange products are formed in abundance alongside radiolysis products. We discuss how our quantitative and qualitative data can be used to interpret the D/H ratios of aliphatic compounds observed throughout space.

*Unified Astronomy Thesaurus concepts:* Ice destruction (2091); Cosmic rays (329); Laboratory astrophysics (2004); Interstellar clouds (834); Solar system (1528)

## 1. Introduction

Deuterium fractionation in interstellar ices is enhanced in low-temperature, dense molecular cores where CO has substantially frozen out (Bacmann et al. 2003), and thus the detection of D-enriched molecules in meteorites (Pizzarello & Huang 2005; Pizzarello & Yarnes 2018) and comets (Balsiger et al. 1995; Bockelée-Morvan et al. 1998; Altwegg et al. 2015) indicate low-temperature interstellar inheritance primarily from ices. One of the main origins of deuterium fractionation is the reaction, $H_3^+ + HD \rightleftharpoons H_2D^+ + H_2$, where the forward reaction dominates at temperatures below 30 K (Smith et al. 1982; Millar et al. 1989). $H_2D^+$ dissociatively recombines (with an electron) to create H- and D-atoms (Larsson et al. 1996); however, this is mitigated when CO is present, as CO readily reacts with $H_3^+$ (Oka 2006). Hence, once CO has frozen out ($T \approx 10$ K), D-atoms are readily produced and contribute along with H-atoms to the formation of icy organics by reaction with solid CO (Hudson & Moore 1999).

There are variations in the D/H ratios in molecular species found in protostellar cores (Cleeves et al. 2016), comets (Cleeves et al. 2016; Altwegg et al. 2019), and meteorites (Remusat et al. 2006; Cleeves et al. 2016). Although the aforementioned low-temperature reactions are largely thought to be responsible for the observed variations in the D/H levels, there are other less well studied processes that can also be influential. An example of such a process is radiation-induced D/H exchange, in which the radiation can come in the form of solar wind particles, cosmic rays, which can penetrate dense molecular cores, secondary electrons, and/or UV photons (Moore et al. 2001; Herbst & Van Dishoeck 2009).

Experimental efforts have shown that the photolysis or the radiolysis of low-temperature ices readily causes D/H exchange between organics and water (Bernstein et al. 1999; Sandford et al. 2000, 2001; Weber et al. 2009). However, the exchange kinetics were not quantified. Thus, the focus of this study is to obtain the D/H exchange rate constants between organics and water in irradiated ices, and specifically of the conversions of C–H → C–D and C–D → C–H.

This work evaluates the rates of radiation-induced D/H exchange in water ice matrices containing organic molecules: $CH_4 + D_2O$ and $CD_4 + H_2O$, and $C_2H_6 + D_2O$ and $C_2D_6 + H_2O$. Experiments involving $C_2D_2 + H_2O$, $C_2D_4 + H_2O$, and $C_2H_4 + D_2O$ were performed but did not yield conclusive results. It should be noted that for each reactant hydrocarbon examined in this study, substitution of any H or D yields only a single isotopologue (e.g., singly substituted product molecules such as $CH_3D$ or $CD_3H$ only possess single isomers, respectively), simplifying the analysis. Solid $CH_4$ is one of the more abundant interstellar ices that should be simultaneously formed with water (Öberg et al. 2008; Boogert et al. 2015). It has also been detected in comets (Mumma et al. 1996; Gibb et al. 2003), planetary atmospheres (Formisano et al. 2004; Swain et al. 2008), and meteorites (Blamey et al. 2015). Acetylene has been detected in comets (Brooke et al. 1996; Weaver et al. 1997; Magee-Sauer et al. 2002; Mumma et al. 2003) and in the atmospheres of Jupiter and Saturn (Noll et al. 1986). It coexists with $C_2H_6$ in such environments, consistent with the results of laboratory experiments that indicate $C_2H_2$ is a parent molecule of $C_2H_6$ (Moore & Hudson 1998; Kobayashi et al. 2017). $C_2H_4$ has only recently been detected in a comet (Russo et al. 2022), but has long been detected along with other

Table 1
Values from the Literature Used to Determine Ice Mixture Ratios

| Molecule (Brand and Purity) | $n$ | $\rho$ [a] g cm$^{-3}$ | References for $n$ and $\rho$ |
|---|---|---|---|
| D$_2$O (Aldrich 99.96% D) | * | 1.04 | ... |
| CH$_4$ (Matheson Ultra High Purity) | 1.30 | 0.47 | Luna et al. (2012) |
| CH$_3$D (Matheson 99% D) | * | 0.50 | ... |
| C$_2$H$_4$ (Matheson 99.99%) | 1.45 | 0.56 | Satorre et al. (2017) |
| C$_2$H$_3$D (Cambridge Isotope Laboratories 98% D) | * | 0.78 | ... |
| C$_2$H$_6$ (Air Products CP Grade) | 1.31 | 0.44 | Satorre et al. (2017) |
| C$_2$H$_5$D (Cambridge Isotope Laboratories D5 98%) | * | 0.74 | ... |
| H$_2$O (Fisher Chemical HPLC Grade) | 1.31 | 0.94 | Weast et al. (1984), Narten et al. (1976) |
| CD$_4$ (MSD Isotopes 99.2% D) | * | 0.59 | ... |
| CD$_3$H (Cambridge Isotope Laboratories 98% D) | * | 0.56 | ... |
| C$_2$D$_2$ (Cambridge Isotope Laboratories 99% D) | * | 0.82 | ... |
| C$_2$D$_4$ (Cambridge Isotope Laboratories 98% D) | * | 0.86 | ... |
| C$_2$D$_6$ (MSD Isotopes 99.8% D) | * | 0.86 | ... |
| C$_2$D$_5$H (Cambridge Isotope Laboratories) | * | 0.84 | ... |

**Note.** * denotes that $n$ is the value for the nondeuterated form. (a) $\rho$ for all deuterated species is estimated by multiplying $\rho$ of the nondeuterated species (obtained from the literature) by $M_{\text{deuterated}}/M_{\text{non-deuterated}}$.

hydrocarbons in Neptune (Schulz et al. 1999), Jupiter and Saturn (Bézard et al. 2001), and Titan (Roe et al. 2004).

## 2. Experimental Methods

All experiments were performed with a high-vacuum chamber optimized for proton irradiation experiments, with IR spectra measured in transmission. The chamber typically had a pressure of $1-2 \times 10^{-7}$ Torr just prior to deposition. The apparatus design was similar to what was described by Hudson & Ferrante (2020), and therefore only a few additional details relevant to this study are mentioned here. Notably, the vacuum chamber was mounted to the end of the beam line from a Van de Graaff accelerator. For all experiments, a 25 mm diameter NaCl substrate was cooled to 20 K before sample growth. A brass ring surrounded the NaCl substrate to measure the radiation dose. Spectra of the ice before and after proton irradiation were collected by a Nicolet iS50 spectrometer. A spectral range of 5500–500 cm$^{-1}$ was used with 256 scans averaged per spectrum at a resolution of 4 cm$^{-1}$.

Ice growth rates were monitored by interference fringes of a 670 nm laser, which we used to determine the desired water: organic ice ratio and compute the column density of the organic compounds embedded in the ice. The deposition rates of water and organics were controlled by leak valves, which were calibrated over a large range of settings for water. The deposition rate of the organic required to produce the desired water:organic ice ratio was determined using the equation

$$\frac{\text{water deposition rate}}{\text{organic deposition rate}} = \frac{N_{\text{water}}}{N_{\text{organic}}} \times \frac{\rho_{\text{organic}}}{\rho_{\text{water}}} \times \frac{M_{\text{water}}}{M_{\text{organic}}} \times \frac{n_{670,\text{water}}}{n_{670,\text{organic}}}, \quad (1)$$

where $N$ = number of molecules (or moles), $\rho$ = density (g cm$^{-3}$), $M$ = molar mass (g mol$^{-1}$), and $n$ = refractive index at 670 nm. To quantify the amount of the reactant destroyed and product formed, the column densities had to be determined. To obtain the column densities, mixtures of 1:25 of organic reactant:water and organic product:water were created (each done twice). A 1:25 ratio was chosen in order to obtain a more accurate band area (i.e., a larger band area so that the uncertainty in the integration was a smaller fraction of the band's area). Approximately 12 fringes were measured for each mixture and were used to determine the total ice thickness and subsequently the column density using the equations

$$h = \frac{N_{\text{fr}}\lambda}{2\sqrt{n^2 - \sin^2\theta}} \quad (2)$$

$$\text{Column density(water–rich ice)} = \frac{h\rho_{\text{water}}N_A}{M_{\text{water}}}, \quad (3)$$

where $h$ = thickness of entire ice, $N_{\text{fr}}$ = number of fringes, $\lambda$ = wavelength of He–Ne laser (670 nm), and $\theta$ is the incident angle that is in between the laser beam and the plane of the substrate ($\approx 0°$) (Heavens 2011) in Equation (2). In Equation (3), $N_A$ is Avogadro's constant. It is assumed that most of the ice is water, and the column density of the organic is determined by multiplying the column density (Equation (3)) by the fraction that represents the amount of the organic in the water-rich ice. The integrated absorbance versus the column density of the organic for every four fringes was plotted to obtain the slope, or the band strength (Hudson & Ferrante 2020). This band strength value was then used to determine the column densities of the formed product in the D/H exchange experiments. Most reactant and product bands could be easily integrated from the uncorrected IR spectra. In the cases of C$_2$H$_5$D and C$_2$D$_5$H, the bands of interest were still substantially overlapping with bands from the solvent. To counteract this, we subtracted the IR spectra of the initial unirradiated ices from each experiment from all subsequent spectra in their respective series in order to improve the baseline. The bands of each consumed reactant (except that of C$_2$H$_6$) overlapped with the bands of a product. Therefore, in order to determine the column density of the consumed reactant, the column density of the product was used to determine how much of the target band area should be attributed to the product, and then was subtracted out. Table 1 lists the values used to calculate the ice mixture ratios, and

**Table 2**
Measured band Strengths at 20 K

| Molecule | Peak Position cm$^{-1}$ | A[a] cm molecule$^{-1}$ |
|---|---|---|
| $CH_4$ | 1303 | $2.58 \times 10^{-18}$ |
| $CH_3D$ | 1302[b] | $6.93 \times 10^{-19}$ |
| $CH_3D$ | 1154 | $1.63 \times 10^{-18}$ |
| $C_2H_6$ | 1373 | $3.25 \times 10^{-19}$ |
| $C_2H_5D$ | 1306 | $2.29 \times 10^{-19}$ |
| $CD_4$ | 993 | $1.16 \times 10^{-18}$ |
| $CD_3H$ | 1000[b] | $4.26 \times 10^{-19}$ |
| $CD_3H$ | 1286 | $9.23 \times 10^{-19}$ |
| $C_2D_6$ | 1071 | $1.34 \times 10^{-18}$ |
| $C_2D_5H$ | 1068[b] | $8.91 \times 10^{-19}$ |
| $C_2D_5H$ | 1301 | $8.54 \times 10^{-19}$ |

**Note.**
[a] In a 1:25 organic:water ice matrix. (b) Product band area overlaps with reactant band area.

Table 2 lists the band strengths measured for each organic that was quantitatively evaluated.

For the D/H exchange radiolysis experiments (each performed three times), a 1:50 organic:water ice ratio was chosen in order to reasonably isolate the organics from each other while still maintaining readily measurable IR bands for the organic reactants and products. Water and the selected organic were deposited simultaneously through separate leak valves. Ices were grown to a thickness of ~3 μm prior to irradiation.

To determine the dose absorbed by each ice, calculations were performed using the software package, Stopping and Range of Ions in Matter (Ziegler et al. 2010). The energy of the incident protons was ~0.9 MeV (Loeffler & Hudson 2018) at a current of $0.5 \times 10^{-7}$ A. The total integrated current was converted to fluence ($F$), in units of p$^+$ cm$^{-2}$. This value, the proton stopping power ($S$; eV cm$^2$ g$^{-1}$ p$^{+-1}$), and a factor of $1.602 \times 10^{-22}$ to convert units of dose from eV g$^{-1}$ to MGy, were used to determine the absorbed doses in units of MGy by the equation

$$\text{Dose (MGy)} = SF \times (1.602 \times 10^{-22}), \quad (4)$$

where a stopping power of $2.685 \times 10^8$ eV cm$^2$ g$^{-1}$ p$^{+-1}$ was used for all dose calculations.

### 3. Results

The D/H exchange kinetics was approximated as a system of parallel first-order reactions

$$A \xrightarrow{k_1} B \quad (5a)$$

$$A \xrightarrow{k_2} P, \quad (5b)$$

where A is the starting compound, B is the singly deuterated or protiated product, and P represents all other products made from A. The relevant rate constants are $k_1$ and $k_2$ (MGy$^{-1}$). The kinetics system used to interpret the data is simplified since the reactions are irreversible and because the kinetics of the subsequent destruction of B are not explicitly considered. These approximations are reasonable at relatively low radiation doses before reverse reactions or multiple D/H exchange reactions within individual organic molecules start to dominate the kinetics. To determine the rate constants, the integrated rate equations used for A and B (Frost & Pearson 1953) were

$$\frac{[A]}{[A]_0} = e^{-(k_1+k_2)D} \quad (6)$$

$$\frac{[B]}{[A]_0} = \frac{k_1}{k_1 + k_2}(1 - e^{-(k_1+k_2)D}) + c_1, \quad (7)$$

where $[A]_0$ is the initial column density of A, D is the dose (MGy), $c_1$ is a constant related to the presence of the product in the starting ice ($\frac{[B]_0}{[A]_0}$). To extract $k_1$, $k_2$, and $c_1$, a nonlinear least-squares routine was used to simultaneously fit Equations (6) and (7) to the formation and destruction data, and the resulting values are in Table 3.

Figure 1 shows examples of the IR spectra collected at 20 K that were used to determine the column densities of the organic reactant and singly deuterated/protiated product needed for Equations (6) and (7). Spectra from the $CH_4 + D_2O$ and $CD_4 + H_2O$ experiments are shown. For each panel, the change in the IR feature as a function of the radiolytic dose is displayed, in which the IR peaks of the reactants ($CH_4$ and $CD_4$) are decreasing, and the corresponding IR peaks of the products ($CH_3D$ and $CD_3H$, respectively) are increasing. A radiolytic dose of up to 6.5 MGy was chosen to follow the D/H exchange before the initial reactant had largely been consumed and before individual molecules with multiple D/H substitutions became prevalent as these factors will complicate the overall reaction scheme and complexity of the IR spectrum (e.g., multiple organic isotopologues) at larger doses, which will have an impact on the apparent rate constants.

Figure 2 shows the fitted data reflecting the radiolytic formation of $CH_3D$ and destruction of $CH_4$ from $CH_4 + D_2O$ (top left), formation of $CD_3H$ and destruction of $CD_4$ from $CD_4 + H_2O$ (top right), formation of $C_2H_5D$ and destruction of $C_2H_6$ from $C_2H_6 + D_2O$ (bottom left), and formation of $C_2D_5H$ and destruction of $C_2D_6$ from $C_2D_6 + H_2O$ (bottom right). Each data point represents the column density normalized to the maximum column density of the reactant as a function of the dose. Each experiment was performed three times, where the displayed rate constants are the average values for the three experiments, and sigma ($\sigma$) is the standard deviation. Note that the normalization has no impact on the value of the rate constants obtained.

**Table 3**
Radiation-induced D/H Exchange Averaged Rate Constants for Each Singly Deuterated/Protiated Organic Product and c Values

| Ice Mixture 1:50 | Organic Product | $k_1$ MGy$^{-1}$ | $k_2$ MGy$^{-1}$ | $c_1$ |
|---|---|---|---|---|
| $CH_4 + D_2O$ | $CH_3D$ | $0.034 \pm 0.003$ | $0.033 \pm 0.002$ | $0.042 \pm 0.003$ |
| $C_2H_6 + D_2O$ | $C_2H_5D$ | $0.056 \pm 0.002$ | $0.053 \pm 0.008$ | $0.019 \pm 0.014$ |
| $CD_4 + H_2O$ | $CD_3H$ | $0.022 \pm 0.002$ | $0.050 \pm 0.004$ | $0.026 \pm <0.001$ |
| $C_2D_6 + H_2O$ | $C_2D_5H$ | $0.022 \pm 0.003$ | $0.044 \pm 0.003$ | $0.003 \pm 0.002$ |

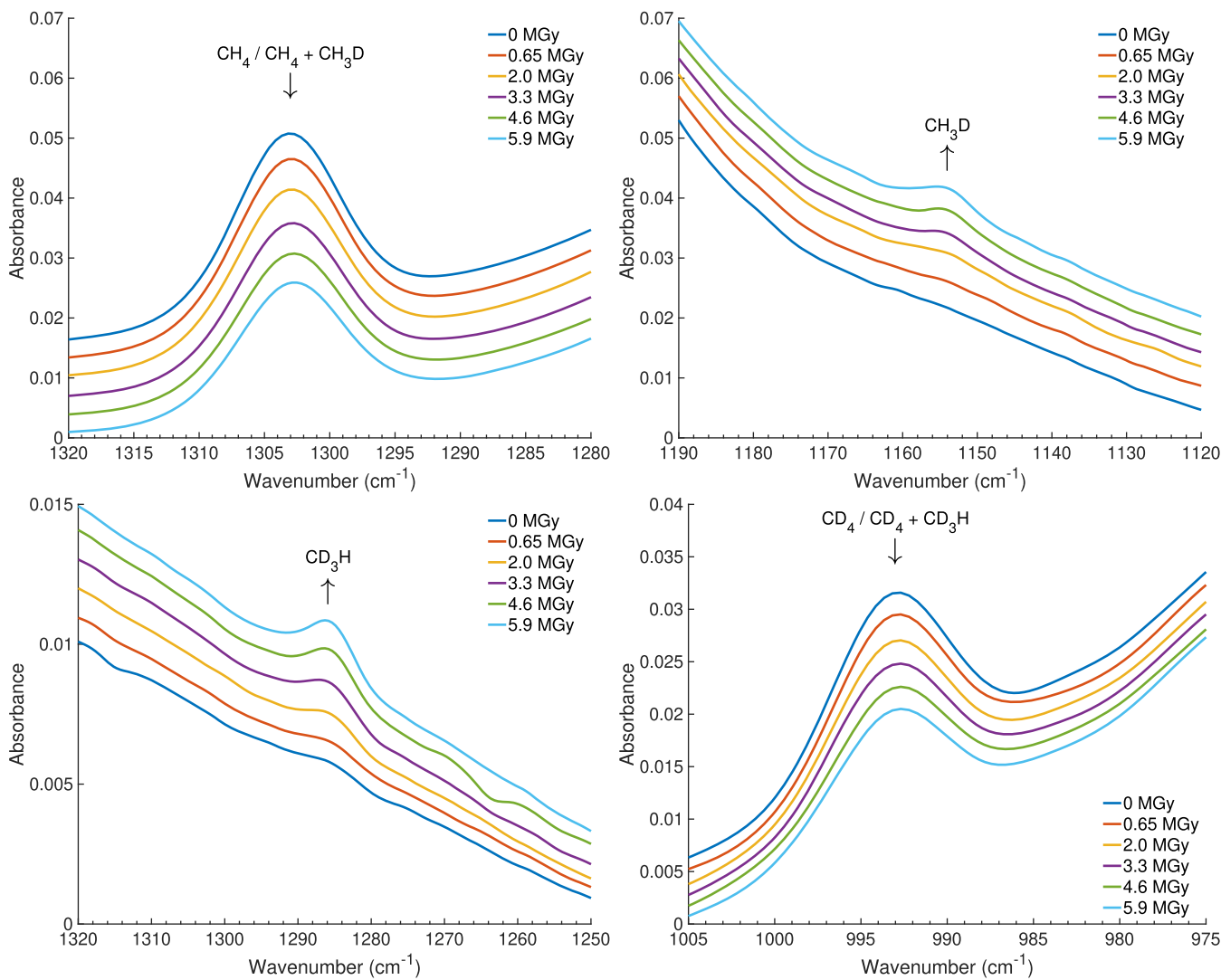

**Figure 1.** The IR spectra of $CH_4 + D_2O$ (top) and $CD_4 + H_2O$ (bottom) 1:50 mixtures at 20 K before and after exposure to irradiation. Arrows identify peaks and indicate the direction of increasing dose. Note that after irradiation, the bands marked as $CH_4$ and $CD_4$ cannot be exclusively attributed to these molecules and also contain some absorbance from $CH_3D$ and $CD_3H$.

In order to demonstrate that the organic reactants would be ultimately consumed if irradiated indefinitely, we conducted high-dose experiments (e.g., Figure 3) that reached a final dose of up to ~329 MGy. By the end of the experiment shown in Figure 3, the $CD_4$ associated band (993 cm$^{-1}$) was mostly gone (less than 8% of its original intensity) and the position of the center of the band had shifted (to 989 cm$^{-1}$), suggesting that what remained might at least in part be attributed to a spectrally overlapping radiolysis product. Furthermore, the intensity of the band continued to decrease with dose, suggesting that it will continue to approach zero. Note that we make no attempt to fit the kinetics of these long experiments for the previously explained reasons.

Although we performed similar radiolysis experiments with $C_2D_4 + H_2O$, $C_2H_4 + D_2O$, and $C_2D_2 + H_2O$ (all 1:50), we were unable to accurately measure the growth of a singly deuterated or protiated product. In these experiments, we were unable to locate any bands clearly associated with the singly deuterated or protiated forms of the molecule that could be cleanly and reliably integrated in a way that gave us confidence in the findings. Consequently, we cannot report rate constants associated with D/H exchange for these compounds.

## 4. Discussion

Overall, we find that the rate constants ($k_1$) for reactions involving H-to-D exchange were greater than those for reactions involving D-to-H exchange. This is consistent with the C–D bond having a lower zero-point energy than the C–H bond. Additionally, Table 3 shows that the rate constant for the conversion of $CH_4$ to $CH_3D$ was lower than that of $C_2H_6$ to $C_2H_5D$. This is consistent with the lower C–H bond energy in ethane in comparison to that in methane (Darwent 1970). Notably, the rate constants for the conversion of $CD_4$ to $CD_3H$ and $C_2D_6$ to $C_2D_5H$ were identical, though the reason for this is not clear.

For both the $D_2O + CH_4$ and $D_2O + C_2H_6$ experiments, $k_1$ and $k_2$ were essentially equivalent, meaning that $CH_3D$ or $C_2H_5D$, respectively, were produced in these experiments at a similar rate to all other radiolysis products combined. For the $H_2O + CD_4$ and $H_2O + C_2D_6$ experiments, $k_2$ significantly exceeded $k_1$. The ratios of the rate constants $k_1$ and $k_2$ provide the relative efficiency of the radiolytic H-to-D exchange versus the radiolytic destruction of the compound. Our results demonstrate that radiation-induced H-to-D exchange in fully protiated methane and ethane are of similar efficiency to the overall radiolytic destruction of these compounds. In contrast, radiation-induced D-to-H exchange in fully deuterated methane and ethane is still significant, but is less efficient than the overall radiolytic destruction of these compounds.

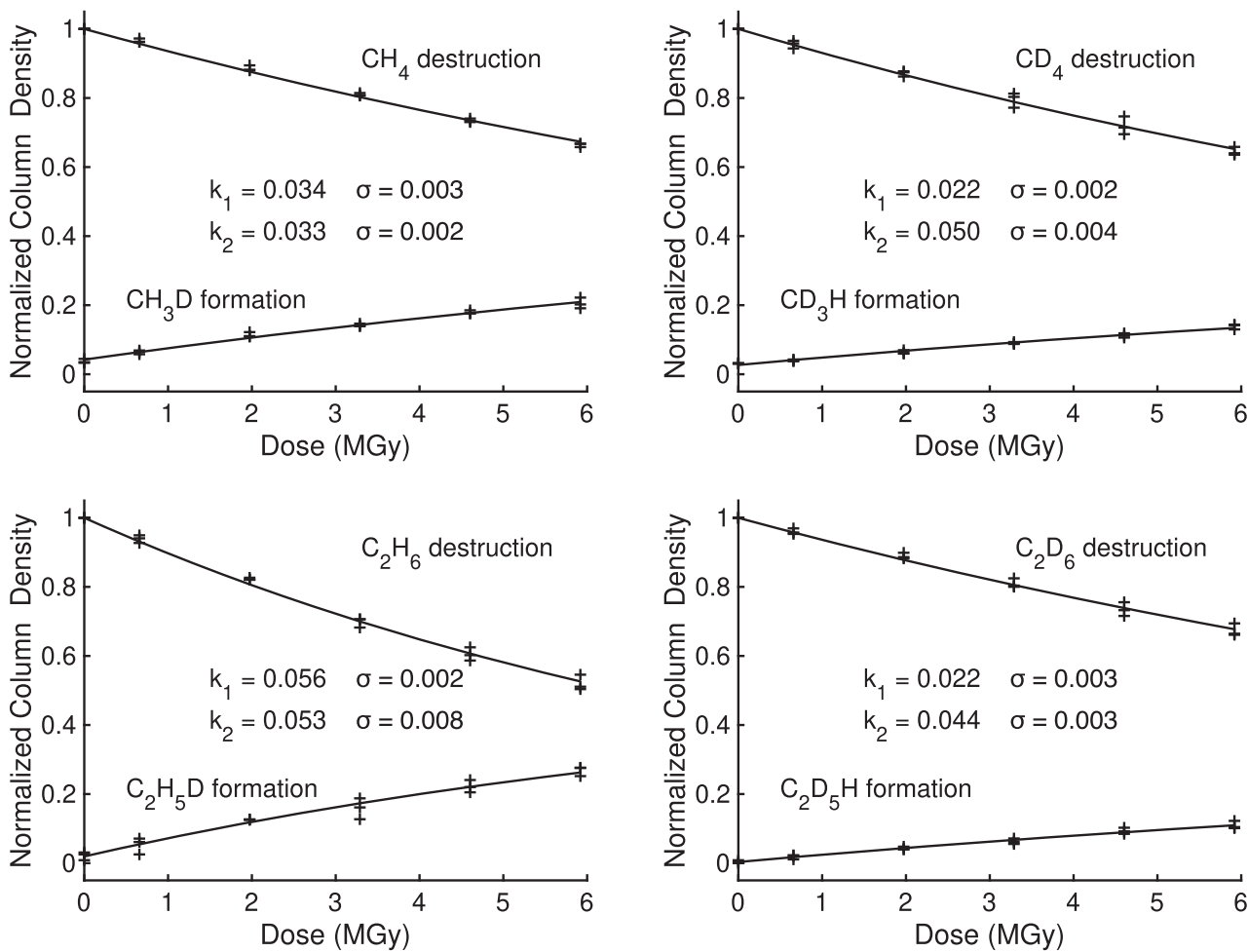

**Figure 2.** Destruction data of each fully protiated or deuterated molecule that were fit using Equation (6), accompanied with the resulting formation data of each singly deuterated or protiated molecule that were fit using Equation (7). Each experiment was performed three times, hence every *x*-axis data point has three *y*-axis data points. Each *y*-axis data point, for reactants and products, represents the column density normalized to the maximum column density of the reactant as a function of the dose. $k_1$ and $k_2$ are the rate constants at 20 K averaged over three experiments, and $\sigma$ is the standard deviation of $k_1$ and $k_2$ between three experiments.

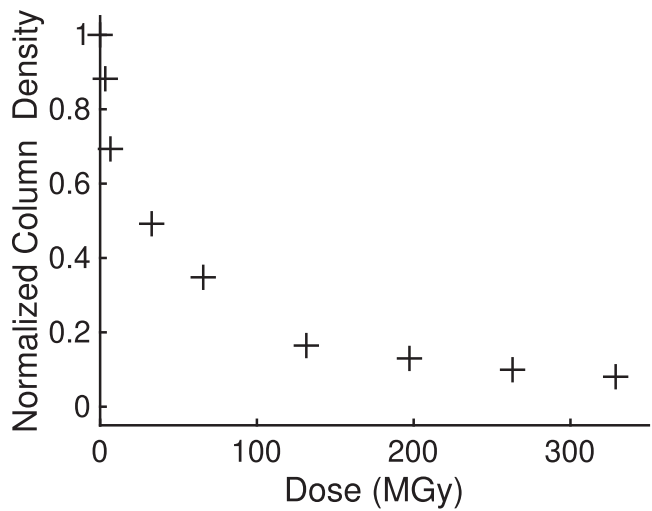

**Figure 3.** Radiolytic destruction data for $CD_4$ in $H_2O$ (1:50) at 20 K based on the 993 cm$^{-1}$ band that has been normalized to the initial column density. Note that at the end of this experiment, less than 8% of the original intensity of the 993 cm$^{-1}$ band remained and the center of the band had shifted to 989 cm$^{-1}$, suggesting that at least part of what remains may be attributed to a spectrally overlapping radiolysis product.

The observed D/H exchanges can be interpreted solely in terms of conventional radiation-chemical processes. In an irradiated 50:1 water:organic ice, the bulk of the energy absorbed from the incident radiation will be by water molecules, either $H_2O$ or $D_2O$. For a $D_2O + CH_4$ ice, the incident protons will eject electrons from water molecules, and in turn those electrons will travel through the ice creating a trail of ionizations and excitations that result in both charged and neutral, open- and closed-shell products. See the reactions below.

$$D_2O \rightarrow D_2O^+ + e^- \tag{8a}$$

$$D_2O^+ + D_2O \rightarrow OD + D_3O^+ \tag{8b}$$

$$D_2O \rightarrow D_2O^+ + e^- \rightarrow [D_2O]^* \rightarrow D + OD. \tag{8c}$$

Molecular products will include $D_2$, $D_2O_2$, and eventually $O_2$. The subsequent changes in $CH_4$ will result from H-atom abstraction to make methyl ($CH_3$) radicals followed by radical–radical coupling, so that both bond breakage and bond formation take place as follows:

$$OD + CH_4 \rightarrow HOD + CH_3 \tag{9a}$$

$$D + CH_3 \rightarrow CH_3D. \tag{9b}$$

However, in the case of $H_2O$-rich ices the abstraction reaction will be slower due to the necessity of abstracting a D-atom from the hydrocarbon, the C–D bond being stronger that the C–H bond. This will result in a smaller rate constant for the D-to-H exchange, as observed. Similar considerations apply to irradiated $D_2O + C_2H_6$ and $H_2O + C_2D_6$ ices.

In contrast, the radiation chemistry of ices made of water and an unsaturated hydrocarbon, such as $D_2O + C_2H_4$ or $D_2O + C_2H_2$ is much more complex. The radicals formed from $D_2O$ can add across the double or triple bond to form alcohols, acetaldehyde, and saturated hydrocarbons (Hudson & Moore 1997; Moore & Hudson 1998). Acetylene also polymerizes easily. This added complexity, coupled with IR bands of products that suffered overlap with reactant features, prevented accurate measurements of D/H exchange in such ices. The greater C–H bond strengths of $C_2H_4$ and $C_2H_2$, compared to $CH_4$ and $C_2H_6$, also hindered observations of D/H exchange.

## 5. Interstellar to Solar System Implications

To determine how much cosmic rays affect the D/H ratios of simple hydrocarbons in various interstellar and solar system environments, the D/H exchange rate constants need to be incorporated into computational models albeit with at least three caveats. One caveat is that the laboratory ices are two-component ices, and therefore the laboratory D/H exchange rate constants should be taken as upper limits, since extraterrestrial ices will contain various organics that will, for example, compete for D/H exchange with water.

The second caveat is that the D/H exchange rate constants will depend on the D/H ratio in the reactant. For example, when $CD_4$ is irradiated in an $H_2O$ matrix, H from the water can only exchange with a D from the methane. In contrast, if $CD_3H$ were irradiated in an $H_2O$ matrix, H from the water could exchange with either D or H from the methane, though the H-to-H exchange cannot be observed. This will cause the rate for D/H to decrease for statistical reasons and also thermodynamic reasons since breaking the C–H bond is more energetically favorable than breaking the C–D bond. Unfortunately, it is difficult to quantify D/H exchange except for fully deuterated or fully protiated hydrocarbons because of overlap of reactant and product IR bands. The formation of multiple isotopologues, common with most hydrocarbons having more than one carbon, further complicates the analysis.

The third caveat is that our rate constants are for 20 K, a temperature more relevant to interstellar than solar system ices. However, these low-temperature rate constants can be used to understand the D/H ratios of organics across the solar system, as there is evidence that interstellar ices and their isotopic ratios are preserved upon distribution to protostellar systems (Jensen et al. 2021), protoplanetary disks (Booth et al. 2021), meteorites (Piani et al. 2021), comets (Altwegg et al. 2019), and even planetary satellites (Mousis et al. 2009). Thus, such low-temperature rate constants are needed to understand the origins of solar system organics that originated in a molecular cloud, and can be used in models with caution.

It is well established in the literature that the bulk nuclear ice of comets is partially composed of pristine interstellar ice that is protected to an extent by layers of cometary material (Altwegg et al. 2019). Therefore, our results can be directly linked to the D/H signatures of cometary organics and water, particularly since the low temperatures of comets best preserve the initial D/H enhancement from the parent molecular cloud. In the molecular cloud, $H_2O$ is initially formed in low-density regions, whereas HDO and $D_2O$ are preferentially formed at high densities (Furuya et al. 2016, 2017), along with deuterated organics due to chemical processes. As found in this study, such organics should have their D/H ratios further enhanced when irradiated in deuterium enriched ice, and therefore this additional enhancement should be considered when evaluating the D/H ratios of pristine cometary organics (i.e., inherited interstellar organics). Conversely, upon irradiation, the additional reduction in the D/H ratio of organics when surrounded by $H_2O$ ice should also be taken into account when evaluating the isotopic origins of cometary organics.

## 6. Conclusion

Experiments were conducted to quantify the radiation-induced D/H exchange rate constants for fully deuterated methane and ethane embedded in $H_2O$ and for fully protiated methane and ethane embedded in $D_2O$ ice at 20 K. Similar experiments were conducted for acetylene and ethylene, but were unable to measure rate constants for D/H exchange, possibly because of the high reactivity of these species that precluded the accumulation of measurable quantities of their singly deuterated or protiated species. Protiated methane and ethane were found to possess different rate constants for the radiolytic exchange of a single hydrogen for deuterium with the surrounding water ice, as consistent with bond energies from the literature. On the other hand, deuterated methane and ethane were found to possess almost the same rate constants. For protiated methane and ethane, this H-to-D exchange with the surrounding ice is an efficient process that is equal to the production of all other radiolysis products combined. For deuterated methane and ethane, D-to-H exchange is still a relatively efficient process but is not generally favored over the combined synthesis of all other potential radiolysis products. The rate of D/H exchange is likely controlled by the breaking of the C–D or C–H bond and the rate constants were measured to be higher for protiated organics in $D_2O$ than for deuterated organics in $H_2O$ ices, and ascribed to the differences in C–H and C–D bond strengths. These effects should be taken into account when studying the D/H ratios of low-temperature interstellar ices preserved in comets.

This work was supported by the Solar System Workings program, award number 18-SSW18-0027, and by NASA, under award number 80GSFC21M0002. D.Q. acknowledges Geronimo Villanueva for stimulating discussions. We acknowledge Stephen Brown, Eugene Gerashchenko, and Martin Carts for maintenance and operation of the Van de Graaff accelerator in the Radiation Effects Facility at GSFC. We thank Perry Gerakines for helping shape the manuscript. We thank Robert Ferrante (USNA) for the development of the substrate, and the many discussions about the substrate.

### ORCID iDs

Danna Qasim https://orcid.org/0000-0002-3276-4780
Reggie L. Hudson https://orcid.org/0000-0003-0519-9429
Christopher K. Materese https://orcid.org/0000-0003-2146-4288